\documentclass[conference,a4]{IEEEtran}
\usepackage{tikz}
\usepackage{epsf,times,amsmath,color}

\usepackage{epsfig,amsfonts,amsmath,graphicx}
\usepackage{subfigure,graphics,amssymb,amsxtra,color}
\usepackage{flushend}
\usepackage{bm}

\usepackage{color,soul}
\usepackage{setspace}
\usepackage[hang,flushmargin,multiple]{footmisc}
\usepackage{epsfig}
\usepackage{wrapfig}
\usepackage{float}
\usepackage{amsfonts}
\usepackage{cite}
 \usepackage{acronym}
\usepackage{mymaths}

\newtheorem{lemma}{Lemma}
\newtheorem{theorem}{Theorem}

\renewcommand{\M}{\bm{\Phi}}

\renewcommand{\B}{\bm{B}}

\acrodef{BCC}{broadcast channel with confidential messages}
\acrodef{CP}{cyclic prefix}
\acrodef{ZS}{zero pad suffix}
\acrodef{pdf}{probability density function}
\acrodef{iid}{independent and identically distributed}
\acrodef{BER}{bit error rate}
\acrodef{OFDM}{orthogonal frequency division multiplexing}
\acrodef{GSVD}{generalized singular value decomposition}
\acrodef{SVD}{singular value decomposition}
\acrodef{DMT}{discrete multitone}
\acrodef{ISI}{intersymbol interference}
\acrodef{ICI}{interchannel interference}
\acrodef{LOS}{line of sight}
\acrodef{NLOS}{non line of sight}
\acrodef{SNR}{signal-to-noise ratio}
\acrodef{PSNR}{peak signal-to-noise ratio}
\acrodef{SINR}{signal to interference plus noise ratio}
\acrodef{SIR}{signal to interference ratio}
\acrodef{MSE}{mean squared error}
\acrodef{MIMO}{multiple-input multiple-output}
\acrodef{MIMOME}{multiple-input multiple-output multiple-eavesdropper}
\acrodef{FFT}{fast Fourier transform}
\acrodef{IFFT}{inverse fast Fourier transform}
\acrodef{CDF}{cumulative distribution function}
\acrodef{CCDF}{complementary cumulative distribution function}
\acrodef{QAM}{quadrature amplitude modulation}
\acrodef{MMSE}{minimum mean-squared error}
\acrodef{LMMSE}{linear minimum mean square error}
\acrodef{SNR}{signal-to-noise ratio}
\acrodef{i.i.d.}{independent identically distributed}
\acrodef{SVD}{singular value decomposition}
\acrodef{MAP}{maximum a posteriori}
\acrodef{MIMO}{multiple input multiple output}
\acrodef{OFDM}{orthogonal frequency division multiplexing}
\acrodef{CSI}{channel state information}
\acrodef{AWGN}{additive white Gaussian noise}
\acrodef{CDF}{cumulative distribution function}
\acrodef{KKT}{Karush-Kuhn-Tucker}
\acrodef{PDP}{power delay profile}
\acrodef{QPSK}{quadrature phase-shift keying}
\acrodef{CS}{compressed sensing}
\acrodef{GMM}{Gaussian mixture model}
\acrodef{OMP}{orthogonal matching pursuit}
\acrodef{EM}{expectation maximization}
\acrodef{pmf}{probability mass function}
\acrodef{DCT}{discrete cosine transform}

\DeclareMathOperator{\rank}{rank}
\DeclareMathOperator{\tr}{tr}

\title{Achievable Secrecy Rates over MIMOME Gaussian Channels with GMM Signals in Low-Noise Regime}

\author
{\IEEEauthorblockN{Francesco Renna} 
\IEEEauthorblockA{Instituto de Telecomunica\c{c}\~{o}es%\\ Departamento de Ci\^{e}ncia de Computadores
\\ Universidade do Porto%\\ Rua Campo Alegre 1021/1055, 4169-007 Porto, Portugal 
\\e-mail: frarenna@dcc.fc.up.pt}
\and 
\IEEEauthorblockN{Nicola Laurenti, Stefano Tomasin}
\IEEEauthorblockA{Dipartimento di Ingegneria dell'Informazione\\ Universit\`a di Padova%\\ via Gradenigo 6/B, 35131 Padova, Italy 
\\ e-mail: \{nil, tomasin\}@dei.unipd.it}}

\begin{document}

\maketitle

\begin{abstract}
We consider a wiretap \ac{MIMOME} channel, where agent Alice aims at transmitting a secret message to agent Bob, while leaking no information on it to an eavesdropper agent Eve. We assume that Alice has more antennas than both Bob and Eve, and that she has only statistical knowledge of the channel towards Eve. %, and Bob has more antennas than Eve. 
%We investigate a compressed sensing approach, where the secret message determines the distribution of a multivariate \ac{GMM} from which a realization is generated and transmitted over the channel (sensing operation). Assuming that Alice has more antennas than both Bob and Eve, the channel operates a compression of the transmitted signal. 
%
%Finally, Bob and Eve attempt reconstruction of the secret message from their observed samples. 
We focus on the low-noise regime, and assess the secrecy rates that are achievable when  the secret message determines the distribution of a multivariate \ac{GMM} from which a realization is generated and transmitted over the channel.
 In particular, we show that if Eve has fewer antennas than Bob, secret transmission is always possible at low-noise. Moreover, we show that in the low-noise limit the secrecy capacity of our scheme coincides with its unconstrained capacity, by providing a class of covariance matrices that allow to attain such limit without the need of wiretap coding.
 
%The main results of our work are the following:
%a) we devise a system for information-theoretic physical layer secrecy in a compressed sensing wiretap channel, where the transmitted signal has a multivariate \ac{GMM} distribution of order $K$ and the informative message is the index in $\{1, . . . , K\}$ of the Gaussian distribution, 
%b) we prove that in the low-noise (or equivalently, high-\ac{SNR}) limit, the secrecy rate that can be achieved by such a system equals the unconstrained capacity, $\log K$, and 
%c) we derive the \ac{GMM} parameters that maximize the secrecy rate  achievabel by our scheme.
\end{abstract}
\acresetall


\acresetall

\section{Introduction}

The application of information theoretic secrecy principles to widely used communication systems is a rising research topic, in an effort to extend security to the lowest layers. Seminal works from the '70s have established the secrecy capacity of a wiretap channel where agent Alice aims at transmitting a secret message to agent Bob while not revealing any information to an eavesdropper agent Eve \cite{Wyner-75,Leung-Yan-Cheong1978}. Since then, a number of other scenarios have been investigated, including the \ac{BCC} case where multiple receivers  require a common message and possibly a different secret message each \cite{Csizar-78}, the case of secret message transmission over parallel channels \cite{Trappe06,Laurenti14}, fading channels \cite{Tomasin14} and the \ac{MIMOME} case \cite{Liu-2009, Wornell-10,Tomasin13}. In this paper we focus on the \ac{MIMOME} channel which may find many important applications in wireless communication systems where the transmitter and the receivers are equipped with multiple antennas. 
In particular, we consider the case in which Alice has perfect knowledge of the \ac{CSI} of the legitimate channel, whereas she has access only to the statistical description of the channel to Eve. A similar scenario was considered in \cite{He10}, in which Alice and Bob were assumed to deploy more antennas than Eve, and achievable secrecy rates were obtained by using wiretap coding schemes.

In conventional transmission techniques one codeword in a fixed set (shared among all users) is transmitted by Alice, and the randomness is due to the noise introduced by the channel or by possible fading. Here instead we consider the scenario in which the transmitted codeword is randomly generated from a continuous set, the secret message determining only the statistics of the random codeword through a map that is known to all agents. This approach has been proposed in \cite{Reeves11}, where a multiplicative Gaussian wiretap channel is considered. In that case the transmitted message is a vector obtained by multiplying a message vector with entries in $\{0, 1\}$ by a diagonal matrix with independent zero-mean unit-variance Gaussian entries. The message vector is assumed to be sparse (i.e., with many zeros). Assuming that Eve has fewer observations than Bob through a known and fixed channel matrix, it is shown that secrecy transmission is possible and lower and upper bounds to the secrecy capacity are derived, when Alice has knowledge of both channels.

In this paper we consider a wiretap \ac{MIMOME} channel scenario where Alice has only statistical knowledge on the channel to Eve. Moreover, 
 %conditions are known to all agents and Alice adopts a compressed sensing transmission technique. However, 
 differently from \cite{Reeves11} we do not restrict the transmitted message to be taken from a binary sparse distribution but we allow denser discrete signaling. As a result, the transmitted signal is a \ac{GMM} multivariate vector, whose statistics must be estimated by the receiver. Moreover, we carry out our analysis in the finite-dimension regime, that is, we assume the number of antennas at Alice, Bob and Eve to be finite. We then focus on the low-noise regime, characterizing the achievable rate of this scheme in the absence of noise, where secrecy is provided by the different channels between the agents. %The design of the sensing matrices\nota[FR]{I would not say we address the design of the sensing matrix} \nota[NL]{Ok, togliamo la frase?} and the statistics of the message vector are also addressed. 
We tailor the statistics of the message vector to maximize the secrecy rate.
The main results provided by our paper are the following:
\begin{enumerate}
\item we devise a system for information-theoretic physical layer secrecy where the transmitted signal is generated from one of $K$ different Gaussian distributions with indices $\{1, . . . , K\}$ and the informative message is the chosen distribution index;
\item we prove that in the low-noise (or equivalently, high-\ac{SNR}) limit, the secrecy rate that can be achieved by such a system equals the unconstrained capacity%\footnote{$\log$ denotes base-2 logarithms.}
, $\log K$, even when Alice has only statistical knowledge of the channel to Eve, and it can be obtained without resorting to wiretap coding techniques;
\item we derive the \ac{GMM} parameters that maximize the secrecy rate achieved by our scheme.
\end{enumerate}

The rest of the paper is organized as follows. Section~II describes the system model, providing details on the \ac{MIMOME} channel, as well as on the specific transmission procedure. Section III focuses on the low-noise regime and we obtain the main results on the achievable rate that can be obtained with the proposed scheme. Numerical results are presented in Section IV, before conclusions are outlined in Section V.

Throughout the paper, vectors (resp., matrices), both deterministic and random, are denoted by  boldface lowercase (resp., uppercase) Latin or Greek letters, while $\log$ denotes the base 2 logarithm. 

\section{System Model}

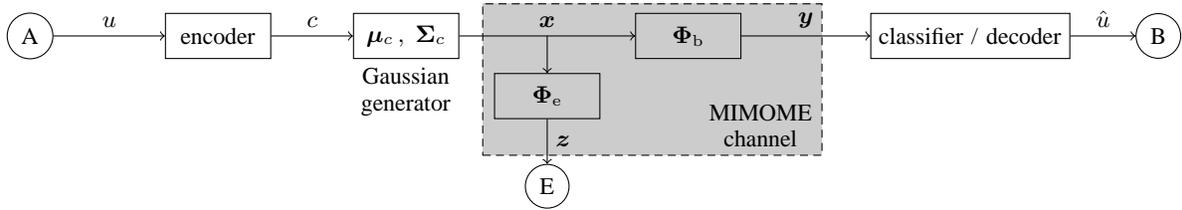
\begin{figure*}
\begin{center}
\begin{small}
\begin{tikzpicture}[x=2.5cm,y=1cm]
%\tikzstyle{every rectangle node}=[text width=1.2cm,minimum width=1.4cm,minimum height=6mm]
\node[rectangle,draw,minimum width=1.4cm,minimum height=6mm](enc) at (0,0) {encoder};
\node[rectangle,draw,minimum width=1.4cm,minimum height=6mm](gau) at (1,0) {$\bm\mu_c\,,\;\bm\Sigma_c$};
\node[text centered,text width=1.3cm,anchor=north](gen) at (gau.south) {Gaussian generator};
\node[rectangle,draw,minimum width=1.4cm,minimum height=6mm](phiB) at (2.5,0) {$\bm\Phi\sub b$};
\draw[->] (gau) -- (phiB) node[midway,above](x){$\bm x$};%\sim\mathcal{N}(\bm{\mu}_c,\bm{\Sigma}_c)$};
\node[rectangle,draw,minimum width=1.4cm,minimum height=6mm](phiE) at (x |- 0,-0.8) {$\bm\Phi\sub e$};
\node[circle,draw](A) at (-1,0) {A};
\node[rectangle,draw,text centered,minimum width=1.4cm,minimum height=6mm](dec) at (4,0) {classifier / decoder};
\node[circle,draw](B) at (5,0) {B};
\node[circle,draw](E) at (x |- 0,-2) {E};
\draw[->] (A) -- (enc) node[midway,above]{$u$};
\draw[->] (enc) -- (gau) node[midway,above]{$c$};
\draw[->] (gau -| phiE) -- (phiE);
\draw[->] (phiE) -- (E) node[midway,right](z){$\bm z$}; 
\draw[->] (phiB) -- (dec) node[midway,above](y){$\bm y$}; 
\draw[->] (dec) -- (B) node[midway,above]{$\hat u$}; 
\draw[densely dashed,fill,fill opacity=0.2] (z.south -| phiE.west) ++ (-1.5mm,0mm) rectangle (y.north east);
\node[text centered,text width=1.4cm,anchor=south east] at (z.south -| y.east){MIMOME channel};
\end{tikzpicture}
\end{small}
\caption{System model. The confidential message $u$ is encoded into the information bearing index $c$.}
\label{fig:sist}
\end{center}
\end{figure*}

We consider a wireless \ac{MIMOME} transmission scenario, as depicted in Fig.~\ref{fig:sist}, in which agent Alice aims at transmitting to agent Bob a secret message $u$ which must be kept secret to a third agent Eve. Alice is equipped with $n$ antennas, while Bob and Eve have $m\sub b$ and $m\sub e$ antennas respectively, with $m\sub b, m\sub e < n$. Between each couple of antennas an \ac{AWGN} flat static channel is available, whose gain does not change for the duration of the entire transmission. At time $t$, Alice transmits a column vector $\bm{x}$ of $n$ symbols on her antennas, and here we assume that $\bm{x}$ has real entries, leaving the extension to complex-valued transmission for future study. The signal vectors received by Bob and Eve have dimension $m\sub b$ and $m\sub e$, respectively, and they can be written as 
\begin{equation}
\begin{array}{r@{\;=\;}l}
\bm{y}&  \bm{\Phi}\sub b \bm{x} + \bm{w}\sub b\\
[0.8mm]  \bm{z}  & \, \bm{\Phi}\sub e \bm{x} + \bm{w}\sub e,
\end{array}
\label{eq:s_model}
\end{equation}
where $\bm{w}\sub b, \bm{w}\sub e \sim \mathcal{N}(\bm{0}, \bm{I}\sigma^2)$ represent \ac{AWGN} noise. Matrices $\bm{\Phi}\sub{b} \in \mathbb{R}^{m\sub b \times n}$ and  $\bm{\Phi}\sub{e} \in \mathbb{R}^{m\sub e \times n}$, represent the \ac{MIMOME} channel. We assume that Alice and Bob know $\bm{\Phi}\sub b$ whereas they have access only to the statistical description of $\bm{\Phi}\sub e$. Eve is assumed to perfectly know both channel matrices.

We consider an average power constraint on the transmitted signal
%, or at least achievable secrecy rates, subject to an average power constraint
\begin{equation}
 \E{\bm{x}\tra \bm{x}} \leq P.
\label{eq:trConsta}
\end{equation}

\subsection{Transmission Technique}

We assume that Alice and Bob agree before transmission on a set of $K$ column vectors of size $n$, $\bm{\mu}_k$, $k=1, \ldots, K$, and $K$ $n\times n$ positive semidefinite matrices $\bm{\Sigma}_k$, $k=1, \ldots, K$. These vectors and matrices are assumed also to be known to Eve. Then, at each transmission Alice encodes by an error correcting code the message $u$ into the the message $c \in \{1, \ldots, K\}$ that is sent by generating vector $\bm{x}$ at random, taken from the multivariate normal distribution $\mathcal{N}(\bm{\mu}_c,\bm{\Sigma}_c)$.

Let $p_k$ be the probability that message $k=1, \ldots, K$ is transmitted. Then the input signal $\bm{x} \in \mathbb{R}^n$ follows a \ac{GMM} distribution with \ac{pdf} 
\begin{equation}
p_{\bm{x}}(\bm{a}) = \sum_{k=1}^K p_k \nu(\bm{a}; \bm{\mu}_k,\bm{\Sigma}_k),
\end{equation}
where 
%\begin{equation}
%\begin{split}
%\nu(\bm{a}; \bm{\mu}_k,\bm{\Sigma}_k) = & \frac{1}{\sqrt{(2\pi)^n\det \bm{\Sigma}_k}}  \\
%& \times \exp{\left(-\frac{1}{2} (\bm{a} -  \bm{\mu}_k)^T \bm{\Sigma}_k^{-1} (\bm{a} -  \bm{\mu}_k)  \right)}
%\end{split}
%\end{equation}
\begin{equation}
\nu(\bm{a}; \bm{\mu}_k,\bm{\Sigma}_k) =  \frac{ \exp{\left(-\frac{1}{2} (\bm{a} -  \bm{\mu}_k)\tra \bm{\Sigma}_k^{-1} (\bm{a} -  \bm{\mu}_k)  \right)}}{\sqrt{(2\pi)^n\det \bm{\Sigma}_k}}  
\end{equation}
is the \ac{pdf} of a multivariate normal distribution. In other terms, the input signals are drawn with probability $p_k$ from the Gaussian distribution $\mathcal{N}(\bm{\mu}_k,\bm{\Sigma}_k)$. Let the rank of the input covariance matrix associated to the $k$-th Gaussian distribution be $s_k=\rank \bm{\Sigma}_k$, and let $s\sub{max} = \max_k s_k$. The discrete random variable $c$ corresponding  to the choice of the particular Gaussian input class has alphabet $\{1, \ldots, K\}$ and \ac{pmf}  $\{p_1, p_2,\ldots, p_K\}$. Note that, in our scenario, the information carried by the transmitted signal is associated with the particular realization of the random variable $c$, rather than the actual value of the vector $\bm{x}$.

% Note that (\ref{eq:s_model}) can seen as the compressed measurements of the \ac{GMM} source signal $\bm{x} \in \mathbb{R}^n$.

Also, the power constraint can be rewritten now as 
\begin{equation}
 \E{\bm{x}\tra \bm{x}} =  \sum_{k=1}^K p_k \tr  (\bm{\Sigma}_k + \bm{\mu}_k\bm{\mu}_k\tra) \leq P.
\label{eq:trConst}
\end{equation}

\subsection{Problem Statement}

Our objective is to determine the \ac{GMM} parameters, that is $\{ p_k \},\{ \bm{\mu}_k\}$ and $\{ \bm{\Sigma}_k \}$, that maximize the achievable secrecy rate. We recall that the supremum on the achievable secrecy rates, i.e., the secrecy capacity, is given in this case by~\cite{Csizar-78}
%\nota{Check if you want $1/n$ here and in the trace constraint}
%the secrecy capacity of the system
\begin{equation}
C\sub s = \max_{(U,c)}  \left[   \mathbb{I}(\bm{y};U)  -  \mathbb{I}(\bm{z};U)   \right],
\label{eq:sec_cap}
\end{equation}
where $U \rightarrow c \rightarrow (\bm{y},\bm{z})$ form a Markov chain and $\mathbb{I}(a;b)$ is the mutual information between $a$ and $b$. In fact, the secrecy capacity (\ref{eq:sec_cap}) is obtained assuming that all agents have perfect \ac{CSI} about all channels. When the channel to Eve is known only statistically by Alice, (\ref{eq:sec_cap}) can be achieved with a given outage probability \cite{Zhou11}. However, we will present in Section~\ref{par:low} a signaling strategy that achieves (\ref{eq:sec_cap}) with $\mathbb{I}(\mathbf{z};c)\to 0$ irrespective of the eavesdropper channel realization.

%\section{High-SNR regime}
%\section{Noiseless case: $\sigma^2 = 0$}

\section{Low-noise regime}
\label{par:low}

We focus our analysis on the low-noise regime, i.e., in the limit $\sigma^2 \to 0$, and we consider the achievable secrecy rates that are obtained by imposing $U=c$ in (\ref{eq:sec_cap}).
%We start considering the case in which $\sigma^2= 0$, and determine a lower bound to (\ref{eq:sec_cap}) similar to that provided in \cite{Reeves11}. We first note that, if we assume
We assume that the number of antennas at Bob are sufficient to guarantee perfect discrimination among signals coming from the different Gaussian classes in the low-noise regime. Namely, we assume
\begin{equation}
m\sub{b} > s\sub{max},
\end{equation}
and assuming also that the shifted range spaces $\mathcal R(\bm\Sigma_k)+\bm\mu_k$ are all distinct, we have that~\cite{Reboredo13}
\begin{equation}
\lim_{\sigma^2 \to 0} P\sub{err}(\sigma^2) = 0,
\end{equation}
where $P\sub{err}$ denotes the misclassification probability associated with the \ac{MAP} classifier that estimates the class from which the input signal $\bm{x}$ was drawn from the observation of the measurement vector $\bm{y}$. Then, by leveraging Fano's inequality~\cite{Cover91}, we can state that
\begin{equation}
\lim_{\sigma^2 \to 0} \mathbb{H}(c|\bm{y})=0, 
\end{equation}
where $\mathbb{H}(\cdot|\cdot)$ denotes the conditional entropy, and, therefore, 
\begin{equation}
\mathbb{I}(\bm{y};c) = \mathbb{H}(c) \leq \log K,
\label{eq:Iyc}
\end{equation}
where $\mathbb{H}(c)$ is the entropy of $c$ and the upper bound in the right hand side is achieved when $p_k={1}/{K}$, for $k=1,\ldots,K$.

%\nota{Remember to underline the fact that the classes are not overlapping}
%\nota[NL]{What does ``overlapping'' mean?}

%\nota{speak about assumptions on the measurements at the eavesdropper}

On the other hand, we assume that the legitimate receiver can leverage an advantage over the eavesdropper in terms of number of antennas, so that $m \sub e \leq s\sub{max} < m \sub b$. Then, consider the information leaked to the eavesdropper, 
%$\mathbb{I}(\bm{z};c) = h(\bm{z}) - h(\bm{z}|c)$, with $h(\cdot)$ denoting the differential entropy. Conditioned on $c = k$, the random vector $\bm{z}$ follows the Gaussian distribution with mean $\bm \Phi\sub e\bm\mu_k$ and covariance $ \bm{\Phi}\sub e \bm{\Sigma}_k \bm{\Phi}\sub e\tra   + \bm{I}\sigma^2$, we can write the conditional differential entropy of $\bm z$ given $c$ as
\begin{equation}
\mathbb{I}(\bm{z};c) = h(\bm{z}) - h(\bm{z}|c),
\end{equation}
with $h(\cdot)$ denoting the differential entropy. Conditioned on $c = k$, the random vector $\bm{z}$ follows the Gaussian distribution with mean $\bm \Phi\sub e\bm\mu_k$ and covariance $ \bm{\Phi}\sub e \bm{\Sigma}_k \bm{\Phi}\sub e\tra   + \bm{I}\sigma^2$ and we can write the conditional differential entropy of $\bm z$ given $c$ as
\begin{IEEEeqnarray}{rCl}
h(\bm{z}| c) & = & \sum_{k=1}^K p_k \frac{1}{2} \log \left[   (2 \pi e )^{m\sub e}  \det \left(   \bm{\Phi}\sub e \bm{\Sigma}_k \bm{\Phi}\sub e\tra   + \bm{I}\sigma^2  \right)  \right]. \IEEEeqnarraynumspace
\end{IEEEeqnarray}

Moreover, note that also the eavesdropper observation $\bm{z}$ follows a \ac{GMM} distribution
\begin{equation}
p_{\bm{z}}(\bm b) = \sum_{k=1}^K p_k \nu(\bm{b}; \bm{\Phi}\sub e \bm{\mu}_k, \bm{\Phi}\sub e \bm{\Sigma}_k \bm{\Phi}\sub e\tra  + \bm{I}\sigma^2).
\end{equation}
%To simplify the analysis, consider for the time being the case of zero-mean \ac{GMM} classes, i.e., $\bm{\mu}_k=\bm{0}$. 

Then, we consider the achievable secrecy rate that is obtained by upper bounding the differential entropy of $\bm z$ by that of a multivariate normal distribution with the same mean vector and covariance matrix~\cite{Cover91}
\begin{IEEEeqnarray}{rCl}
\nonumber
\bm{\mu}_{\bm{z}} &=& \E{\bm{z}}= \sum_{k=1}^K p_k \bm{\Phi}\sub e \bm{\mu}_k\\
\nonumber
\bm{\Sigma}_{\bm{z}}& =&  \E{  (\bm{z} - \bm{\mu}_{\bm{z}}) (\bm{z} - \bm{\mu}_{\bm{z}})\tra}\\
\nonumber
& =&  \sum_{k=1}^K p_k \bm{\Phi}\sub e \left(   \bm{\Sigma}_k + \bm{\mu}_k \bm{\mu}_k\tra \right) \bm{\Phi}\sub e\tra \\
& &- \sum_{k,\ell=1}^K p_k p_\ell  \bm{\Phi}\sub e (\bm{\mu}_k\bm{\mu}_\ell\tra) \bm{\Phi}\sub e\tra + \bm{I}\sigma^2
\label{eq:Sigmaz}
\end{IEEEeqnarray}
respectively, that is by writing
\begin{equation}
h(\bm{z}) \leq h\sub{G}\left({\bm{z}} \right) = \frac{1}{2} \log \left[  (2 \pi e )^{m\sub e}  \det \left(\bm{\Sigma}_{\bm{z}} \right)  \right].
\label{eq:maxEn}
\end{equation}
Therefore, a lower bound to the secrecy rate achieved in this scenario is
\begin{IEEEeqnarray}{rCl}
R\sub s (\sigma^2 )  &= & \mathbb{I}(\bm{y};c) - h\sub G (\bm{z}) + h(\bm{z}|c) \\
\nonumber
 & =  & \mathbb{I}(\bm{y};c) -\frac{1}{2} \log \left[   (2 \pi e )^{m\sub e}  \det \left(\bm{\Sigma}_{\bm{z}}\right)  \right] \\
 &  + & \sum_{k=1}^K p_k \frac{1}{2} \log \left[   (2 \pi e )^{m\sub e}  \det \left(   \bm{\Phi}\sub e \bm{\Sigma}_k \bm{\Phi}\sub e\tra + \bm{I}\sigma^2   \right)  \right]. \IEEEeqnarraynumspace
\label{eq:Rs}
\end{IEEEeqnarray}

Our aim now is to determine the parameters of the \ac{GMM} distribution that maximize the low-noise limit of the achievable secrecy rate (\ref{eq:Rs})
\begin{equation}
\begin{split}
R\sub{s}^{\rm LN}  &=  \lim_{\sigma^2 \to 0} R_s (\sigma^2 ) \\
 & =  \mathbb{H}(c) -\lim_{\sigma^2 \to 0 }  \left[  h\sub{G} (\bm{z}) - h(\bm{z}|c)   \right].
 \label{eq:RsLN}
\end{split}
\end{equation}
Observe that, in the low-noise regime, the class means $\bm{\mu}_k$ affect the values of $R\sub{s}^{\rm LN}$ only through the term $h\sub{G}(\bm{z})$.

The following Lemma states that using zero-mean classes maximizes the low-noise achievable secrecy rate.

\begin{lemma}
\label{lem1}
Given the positive semidefinite matrix $\bm{\Sigma}_{\bm{z}}$ in (\ref{eq:Sigmaz}), it holds
\begin{equation}
\log \det (\bm{\Sigma}_{\bm{z}})  \geq \log \det \left( \sum_{k=1}^K p_k \bm{\Phi}\sub e \bm{\Sigma}_k \bm{\Phi}\sub e\tra   + \bm{I}\sigma^2 \right).
\label{eq:lemma1}
\end{equation}
\end{lemma}

\begin{IEEEproof}
Consider the difference matrix
%\begin{equation}
%\begin{split}
%\bm{\Delta}  = &\bm{\Sigma}_{\bm{z}}  - \left(\sum_{k=1}^K p_k \bm{\Phi}\sub e \bm{\Sigma}_k \bm{\Phi}\sub e\tra   + \bm{I}\sigma^2  \right) \\
% = & \sum_{k=1}^K p_k   \bm{\Phi}\sub e \bm{\mu}_k (\bm{\Phi}\sub e \bm{\mu}_k)\tra -   \sum_{k=1}^K p_k \bm{\Phi}\sub e \bm{\mu}_k   \sum_{\ell=1}^K p_\ell \bm{\mu}_\ell\tra \bm{\Phi}\sub e\tra. \IEEEeqnarraynumspace
%\end{split}
%\end{equation}
\begin{IEEEeqnarray}{rCl}
\nonumber
\bm{\Delta} & = &\bm{\Sigma}_{\bm{z}}  - \left(\sum_{k=1}^K p_k \bm{\Phi}\sub e \bm{\Sigma}_k \bm{\Phi}\sub e\tra   + \bm{I}\sigma^2  \right) \\
 &= & \sum_{k=1}^K p_k   \bm{\Phi}\sub e \bm{\mu}_k (\bm{\Phi}\sub e \bm{\mu}_k)\tra -   \sum_{k=1}^K p_k \bm{\Phi}\sub e \bm{\mu}_k   \sum_{\ell=1}^K p_\ell \bm{\mu}_\ell\tra \bm{\Phi}\sub e\tra. \IEEEeqnarraynumspace
\end{IEEEeqnarray}
Note that $\bm{\Delta}$ is the covariance matrix of a discrete random vector taking values in the alphabet $\{ \bm{\Phi}\sub e \bm{\mu}_1,\ldots,\bm{\Phi}\sub e \bm{\mu}_K \}$ with \ac{pmf} $\{p_1, \ldots, p_K\}$, and thus, it is positive semidefinite. Then, by leveraging Weyl's Theorem (see Corollary 4.3.3 in \cite{Horn}), we can conclude that the ordered eigenvalues of $\bm{\Sigma}_{\bm{z}}$ are all greater or equal than the corresponding ordered eigenvalues of the matrix 
on the left hand side of (\ref{eq:lemma1}), thus proving the inequality.
\end{IEEEproof}

On the basis of Lemma~\ref{lem1}, we consider secrecy rates achieved by zero-mean classes, that is, choosing $\bm{\mu}_k=\bm{0}$, $k=1,\ldots,K$. Moreover, on leveraging the convergence properties of differential entropy due to dominated convergence~\cite{Piera08}, we can write the low-noise achievable secrecy rate with zero-mean classes as
\begin{equation}
\begin{split}
R\sub s^{\rm LN} = &   \mathbb{H}(c) -\frac{1}{2} \log   \det \left(\sum_{k=1}^K p_k \bm{\Phi}\sub e \bm{\Sigma}_k \bm{\Phi}\sub e\tra  \right) \\
 & +  \frac{1}{2} \sum_{k=1}^K p_k  \log  \det \left(   \bm{\Phi}\sub e \bm{\Sigma}_k \bm{\Phi}\sub e\tra   \right), \IEEEeqnarraynumspace
\end{split}
\end{equation}
and we can determine the supremum of the secrecy rates achievable in the low-noise regime.

\begin{theorem}
\label{th1}
The low-noise secrecy capacity associated with the \ac{MIMOME} system described in (\ref{eq:s_model}) with \ac{GMM} transmission and discrete input $c\in\{1,\ldots,K\}$ is given by
\begin{equation}
C\sub{s}^{\rm LN} =\lim_{\sigma^2 \to 0} C\sub s = \log K.
\end{equation}
\end{theorem}
\begin{IEEEproof}
The converse part of the proof is trivial, and it is based on the fact that the secrecy capacity is always lower or equal than the capacity without secrecy constraints, that implies $C\sub s \leq \max_c \mathbb{I}(\bm{y};c) \leq \mathbb{H}(c) \leq \log K$.

In order to provide the achievability part of the proof, we observe that there is a sequence of input distributions that achieves the upper bound $\log K$ in the asymptotic low-noise regime. The existence of such a sequence is guaranteed, in our scenario, by the fact that, when $\sigma^2 \to 0$, the upper bound in (\ref{eq:Rs}) to $\mathbb{I}(\bm z; c)$ is a continuous function of the input covariance matrices $\bm{\Sigma}_k$, whereas $\mathbb{I}(\bm y; c)$ is not, since (\ref{eq:Iyc}) holds if %the $\bm\Sigma_k$ are all distinct. 
the range spaces $\mathcal R(\bm\Sigma_k)$ are all distinct.

We now define a class of covariance matrices, for which we will be able to show that $\mathbb{I}(\bm z;c)\to 0$.  For a given $\epsilon \in \mathbb{R}$, consider the $K$-classes, zero-mean, \ac{GMM} vectors obtained by choosing $\bm{\mu}_k=\bm{0}, p_k=1/K$ and covariance matrices
\begin{equation}
\label{eq:Sk}
\bm{\Sigma}_k(\epsilon) =  \frac{P}{m\sub b -1} \bm{W}(\epsilon)^k
\left[
\begin{array}{ccc}
\bm{I}_{m\sub b -1} & \bm{0} \\
\bm{0} & \bm{0}
\end{array}
\right] [\bm{W}(\epsilon)^k]\tra,
\end{equation}
where $\bm{W}(\epsilon) \in \mathbb{R}^{n \times n}$ is an orthogonal matrix obtained according to the following Cayley transform \cite{Golub}:
\begin{equation}
\bm{W}(\epsilon) = (\bm{I} - \bm{A}(\epsilon))  (\bm{I} + \bm{A}(\epsilon))^{-1}
\end{equation}
from the skew-symmetric matrix
\begin{equation}
\bm{A}(\epsilon) = \left[
\begin{array}{ccccccc}
0 & -\epsilon & \cdots & -\epsilon \\
\epsilon & 0 & \cdots & -\epsilon \\
\epsilon & \cdots & \ddots & -\epsilon \\
\epsilon & \cdots & \epsilon & 0 \\
\end{array} 
\right]
\end{equation}
Now, it is straightforward to note that
\begin{equation}
\lim_{\epsilon \to 0} \bm{W}(\epsilon) = \bm{I}, 
\end{equation}
and, therefore, when $\epsilon \to 0$, all the matrices $\bm\Sigma_k$ coincide and the two rightmost terms in (\ref{eq:RsLN}) cancel out. %Then, due to the log-concavity property of the determinant of positive definite matrices \nota{we might need a reference for the log-concavity of the determinant of positive semidefinite matrices.} \nota[NL]{Alternatively we can use the fact that $\log_2\det\bm A = \tr\log_2\bm A$, is there any result on $\log(\lambda\bm A+(1-\lambda)\bm B) - \lambda\log_2\bm A - (1-\lambda)\log_2\bm B$ being positive semidefinite?} \nota[NL]{actually we don't need concavity, just continuity}, we conclude that the supremum of the secrecy rates achieved with this input, in the low-noise regime is given by
Then, due to the continuity of the log-determinant function of positive definite matrices, we conclude that the supremum of the secrecy rates achieved with this class of covariance matrices, in the low-noise regime is given by
\begin{equation}
\sup_{\epsilon } R\sub s^{\rm LN} (\epsilon) = \log K.
\end{equation}
\end{IEEEproof}

%\begin{remark}
It is relevant to observe that the signaling scheme determined by the input covariance matrices in (\ref{eq:Sk}) guarantees that, asymptotically, $\mathbb{I}(\mathbf{z};c) \to 0$ with probability 1 for all the possible realizations of the eavesdropper channel matrix $\mathbf{\Phi}\sub e$. This result has been shown to hold in the low-noise limit  $\sigma^2 \to 0$, thus implying that, for all values $\sigma^2>0$, the mutual information to Eve still asymptotically approaches zero, since adding noise decreases the quality of communication.  Therefore the proposed scheme provides secrecy even without the knowledge of the eavesdropper channel and, most notably, without requiring the use of wiretap codes. In fact, error correcting coding only can be used and secrecy is obtained directly by leveraging the fact that the mutual information at the eavesdropper can be reduced asymptotically to zero by tuning the parameter $\epsilon$. Moreover, in the low-noise regime, such scheme achieves the secrecy capacity $\log K$.
%\end{remark}

\section{Numerical results}

In the previous sections, we have described a transmission strategy which achieves a secrecy rate equal to $\log K$ in the low-noise regime without the need of wiretap coding. In this section, we focus on finite \ac{SNR} values, assessing the \emph{equivocation rate}~\cite{WongWong11} 
\begin{equation}
R\sub e = \left[  \mathbb{I}(\bm y;c) - \mathbb{I}(\bm z;c) \right]^+.
\end{equation}
We consider an error correcting code with rate $R\sub c = \mathbb{I}(\bm y;c)$ and recall that, when the equivocation rate $R\sub e$ is equal to the transmission rate $R\sub c $, then we have perfect secrecy~\cite{Wyner-75}. We set $n=10$, $m\sub b = 6$ and $m\sub e =4$, respectively. 

\begin{figure}
\begin{center}
\includegraphics[width=0.53\textwidth,height=6.5cm]{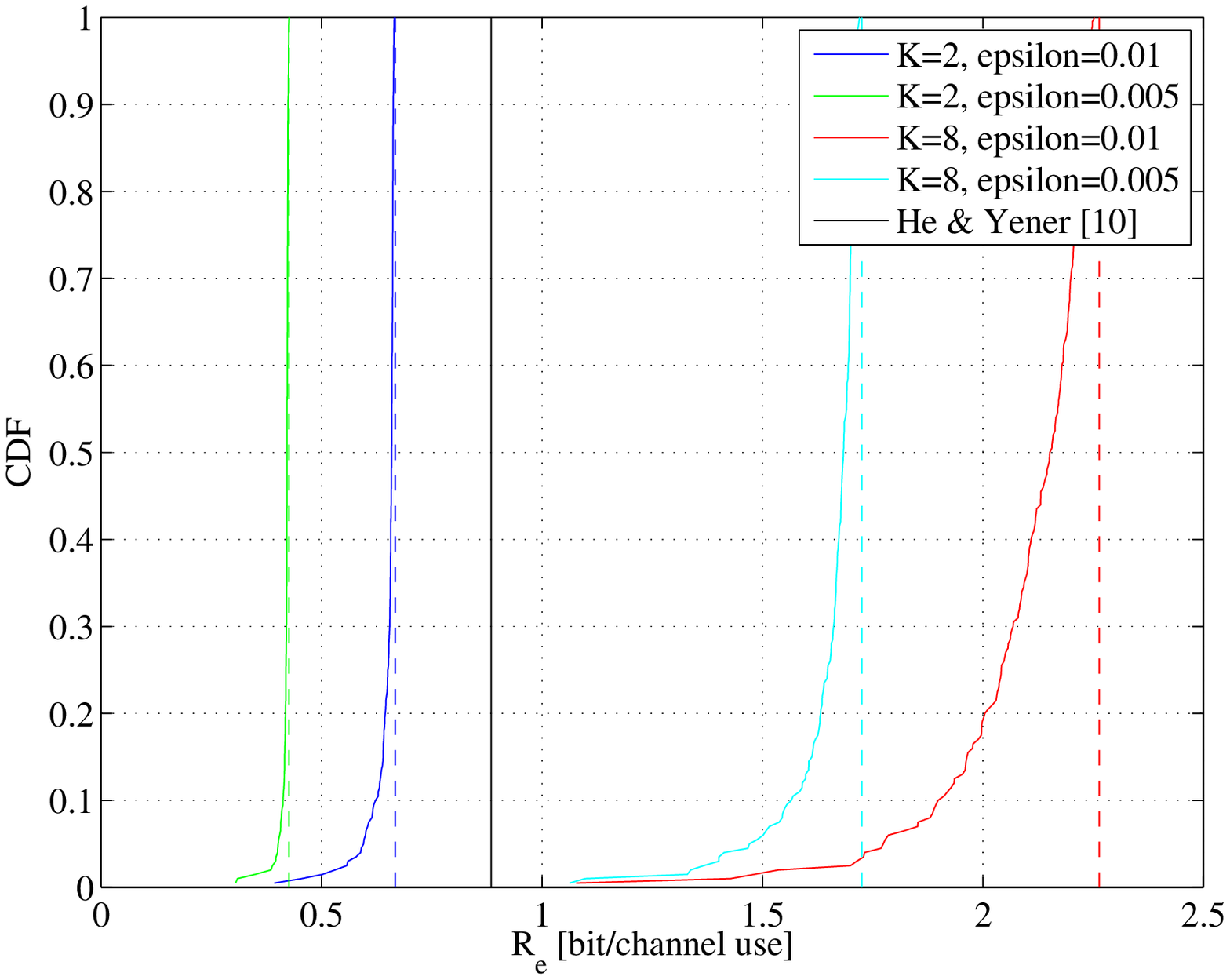}
\caption{CDF of the equivocation rates $R\sub e$ with fixed $\mathbf{\Phi}\sub b$. $\mathrm{SNR}= 35$\,dB, $\epsilon =  0.01, 0.005$. $n=10$, $m\sub b =6$ and $m\sub e =4$, $K=2,8$. The dashed vertical lines represent the transmission rates $R\sub c$.}
\label{fig:Fig1}
\end{center}
\vspace{-6mm}
\end{figure}

Fig.~\ref{fig:Fig1} shows the \ac{CDF} of the equivocation rates obtained when the legitimate channel matrix $\mathbf{\Phi}\sub b$ contains the first $m\sub b$ rows of an $n$-dimensional \ac{DCT} matrix, whereas the eavesdropper channel matrices are randomly generated with \ac{i.i.d.} zero-mean, unit-variance, Gaussian entries. The \ac{SNR} is equal to $35$\,dB, the numbers of classes of the transmitted signals are $K =2$ and $K=8$, and $\epsilon = 0.01$ and $\epsilon = 0.005$.  We also report the value of the secrecy rate $R\sub H$ that is achieved by the wiretap coding scheme described in \cite{He10}. We can notice that, when $K=2$, our scheme provides a much lower equivocation rate than the secrecy rate of \cite{He10}. Moreover, on increasing the number of transmitted classes to $K=8$, higher equivocation rates are achieved than the secrecy rate of \cite{He10}, at the expense of a higher information leakage towards Eve.
%\vspace{-3mm}

\begin{figure}
\begin{center}
\includegraphics[width=0.53\textwidth,height=6.5cm]{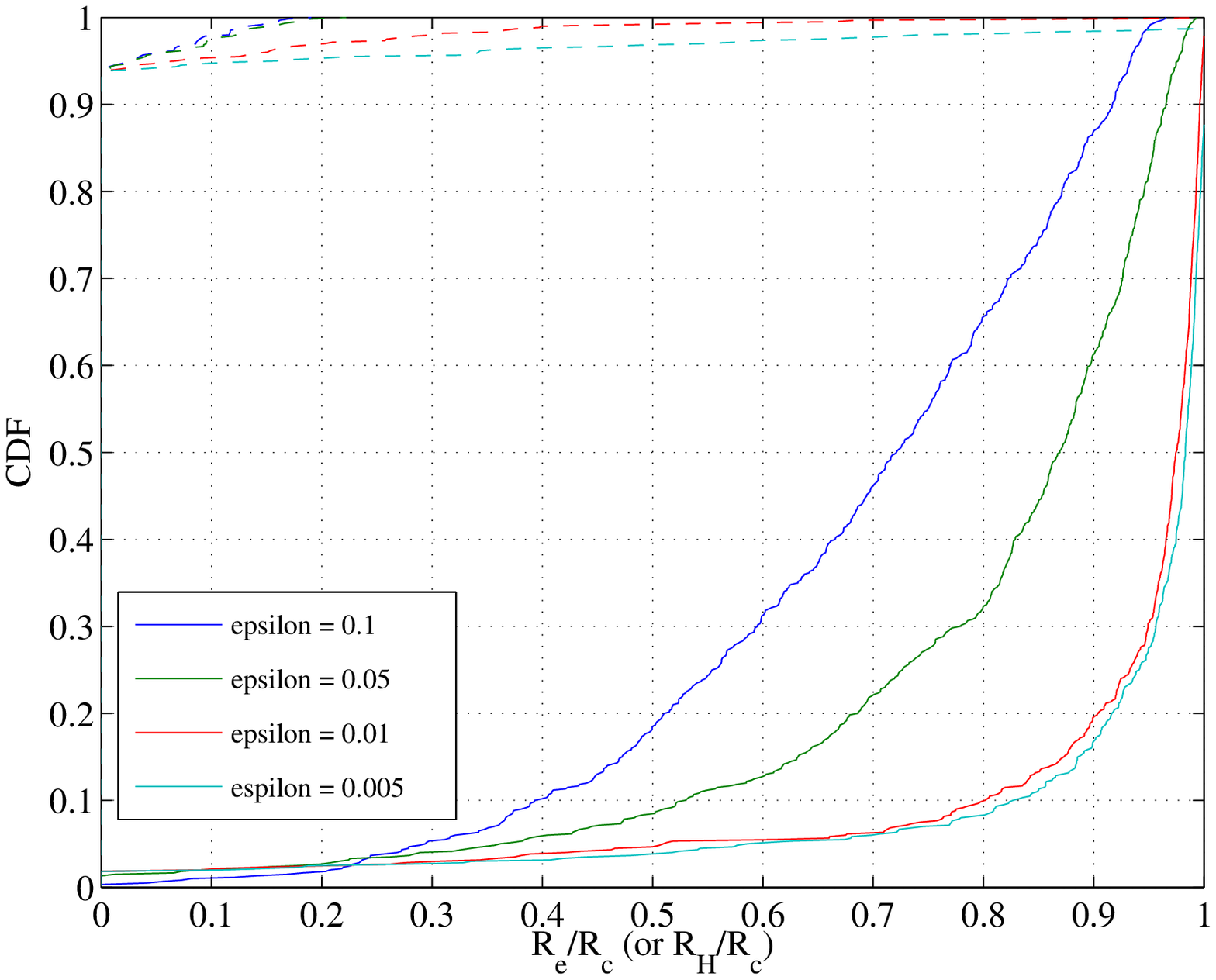}
\caption{CDF of the rate $R\sub e/R\sub c$ (solid lines) and of the ratio $R\sub H/R\sub c$ (dashed lines). $\mathrm{SNR}=25$\,dB. $n=10$, $m\sub b =6$ and $m\sub e =4$, $K=2$ and $\epsilon = 0.1, 0.05, 0.01, 0.005$.}
\label{fig:Fig2}
\end{center}
\vspace{-6mm}
\end{figure}

We then consider the case in which also $\mathbf{\Phi}\sub b$ is generated at random with \ac{i.i.d.}, zero-mean, unit-variance Gaussian entries. Alice is  assumed to know the current realization of the legitimate channel coefficients, and she could arguably optimize the values of $K$ and $\epsilon$ to maximize the equivocation rate under a given constraint on the probability that the leakage to the eavesdropper overcomes a given threshold. Nevertheless, we assess approximately the performance of the system by considering the case in which $K=2$, $\mathrm{SNR}=25$\,dB, and by choosing $\epsilon = 0.1, 0.05, 0.01, 0.005$. Fig.~\ref{fig:Fig2} shows the \ac{CDF} of $R\sub e/R\sub c$, i.e., the secure fraction of the transmitted rate. For comparison, we also report the \ac{CDF} of $R\sub H/R\sub c$, where $R_c$ is still the code rate of our scheme. We observe that, by taking $\epsilon \leq 0.01$,  only the $20 \%$ of the channel realizations correspond to information leakages to Eve that are larger than the $10 \%$ of the transmitted rate. Moreover, for such values of $\epsilon$, the transmission rate guaranteed by our scheme is higher than the secrecy rate of \cite{He10} for the large majority of channel realizations.

\section{Conclusions}

In this paper, we have studied secrecy rates achievable over a \ac{MIMOME} channel when the transmitted signals are drawn from a $K$-classes \ac{GMM} distribution, with the information encoded into the index of the chosen Gaussian class. In particular, we have considered the case in which the legitimate user can deploy more antennas than the eavesdropper but they have only statistical knowledge of the channel to Eve, and we have studied the achievable secrecy rates in the low-noise regime. %Note that this system can be thought as a compressed sensing scenario, in which Eve is assumed to have access to a smaller number of random measurements than Bob.

We have proved that, also when the number of antennas deployed by all the nodes in the network is finite, the low-noise secrecy capacity of this system is given by the unconstrained capacity $\log K$. We have also described a class of \ac{GMM} distributions which achieve the low-noise secrecy capacity by nulling the mutual information at the eavesdropper, thus without the need of wiretap codes. %Finally, we have showcased, through simulation results, the tradeoff that such design needs to strike between reliable decoding and information leakage to the eavesdropper when the noise power $\sigma^2$ is strictly positive.

\section*{Acknowledgements}
This work was supported in part by the MIUR project ESCAPADE (Grant RBFR105NLC) under the ``FIRB-Futuro in Ricerca 2010'' funding program.

\bibliographystyle{IEEEtran}
\bibliography{references_rec}

\begin{thebibliography}{10}
\providecommand{\url}[1]{#1}
\csname url@samestyle\endcsname
\providecommand{\newblock}{\relax}
\providecommand{\bibinfo}[2]{#2}
\providecommand{\BIBentrySTDinterwordspacing}{\spaceskip=0pt\relax}
\providecommand{\BIBentryALTinterwordstretchfactor}{4}
\providecommand{\BIBentryALTinterwordspacing}{\spaceskip=\fontdimen2\font plus
\BIBentryALTinterwordstretchfactor\fontdimen3\font minus
  \fontdimen4\font\relax}
\providecommand{\BIBforeignlanguage}[2]{{%
\expandafter\ifx\csname l@#1\endcsname\relax
\typeout{** WARNING: IEEEtran.bst: No hyphenation pattern has been}%
\typeout{** loaded for the language `#1'. Using the pattern for}%
\typeout{** the default language instead.}%
\else
\language=\csname l@#1\endcsname
\fi
#2}}
\providecommand{\BIBdecl}{\relax}
\BIBdecl

\bibitem{Wyner-75}
A.~Wyner, ``The wiretap channel,'' \emph{Bell System Technical Journal},
  vol.~54, no.~8, pp. 1355--1387, 1975.

\bibitem{Leung-Yan-Cheong1978}
S.~Leung-Yan-Cheong and M.~E. Hellman, ``{The Gaussian wire-tap channel},''
  \emph{IEEE Trans. Inf. Theory}, vol.~24, no.~4, pp. 451--456, Jul. 1978.

\bibitem{Csizar-78}
I.~Csisz\'{a}r and J.~K$\ddot{\mathrm{o}}$rner, ``Broadcast channels with
  confidential messages,'' \emph{IEEE Trans. Inf. Theory}, vol.~24, no.~3, pp.
  339--348, May 1978.

\bibitem{Trappe06}
Z.~Li, R.~Yates, and W.~Trappe, ``Secrecy capacity of independent parallel
  channels,'' in \emph{Allerton Conference in Communication, Control, and
  Computing}, Monticello, IL, 2006.

\bibitem{Laurenti14}
N.~Laurenti, S.~Tomasin, and F.~Renna, ``Resource allocation for secret
  transmissions on parallel {Rayleigh} channels,'' in \emph{IEEE Int. Conf. on
  Commun. (ICC)}, Sidney, Australia, Jun. 2014.

\bibitem{Tomasin14}
S.~Tomasin and N.~Laurenti, ``Secret message transmission by {HARQ} with
  multiple encoding,'' in \emph{IEEE Int. Conf. on Commun. (ICC)}, Sidney,
  Australia, Jun. 2014.

\bibitem{Liu-2009}
T.~Liu and S.~Shamai, ``A note on the secrecy capacity of the multiple-antenna
  wiretap channel,'' \emph{IEEE Trans. Inf. Theory}, vol.~55, pp. 2547--2553,
  Jun. 2009.

\bibitem{Wornell-10}
A.~Khisti and G.~W. Wornell, ``Secure transmission with multiple antennas--part
  {II}: The {MIMOME} wiretap channel,'' \emph{IEEE Trans. Inf. Theory},
  vol.~56, no.~11, pp. 5515--5532, Nov. 2010.

\bibitem{Tomasin13}
S.~Tomasin, ``Resource allocation for secret transmissions over {MIMOME} fading
  channels,'' in \emph{IEEE Global Conference on Commun. (GLOBECOM), Workshop
  on Trusted Communications with Physical Layer Security}, Atlanta, GA, Dic.
  2013.

\bibitem{He10}
X.~He and A.~Yener, ``{MIMO} wiretap channels with arbitrarily varying
  eavesdropper channel states,'' \emph{CoRR}, vol. abs/1007.4801, 2010.

\bibitem{Reeves11}
G.~Reeves, N.~Goela, N.~Milosavljevic, and M.~Gastpar, ``A compressed sensing
  wire-tap channel,'' \emph{CoRR}, vol. abs/1105.2621, 2011.

\bibitem{Zhou11}
X.~Zhou, M.~R. McKay, B.~Maham, and A.~Hj{\o}rungnes, ``{Rethinking the secrecy
  outage formulation: A secure transmission design perspective},'' \emph{IEEE
  Commun. Letters}, vol.~15, no.~3, pp. 302--304, Mar. 2011.

\bibitem{Reboredo13}
H.~Reboredo, F.~Renna, R.~Calderbank, and M.~R.~D. Rodrigues, ``Compressive
  classification,'' in \emph{IEEE Int. Symp. on Inform. Theory (ISIT)},
  Istanbul, Turkey, Jul. 2013.

\bibitem{Cover91}
T.~M. Cover and J.~A. Thomas, \emph{Elements of Information Theory}.\hskip 1em
  plus 0.5em minus 0.4em\relax New York, NY: Wiley, 1991.

\bibitem{Horn}
R.~Horn and C.~Johnson, \emph{Matrix Analysis}.\hskip 1em plus 0.5em minus
  0.4em\relax Cambridge, UK: Cambridge University Press, 1985.

\bibitem{Piera08}
F.~J. Piera and P.~Parada, ``On convergence properties of {Shannon} entropy,''
  \emph{Probl. Inf. Transm.}, vol.~45, pp. 75--94, June 2009.

\bibitem{Golub}
G.~H. Golub and C.~F.~V. Loan, \emph{Matrix Computations}.\hskip 1em plus 0.5em
  minus 0.4em\relax Baltimore: Johns Hopkins University Press, 1996.

\bibitem{WongWong11}
C.~W. Wong, T.~Wong, and J.~Shea, ``Secret-sharing {LDPC} codes for the
  {BPSK}-constrained {G}aussian wiretap channel,'' \emph{IEEE Trans. Inf.
  Forensics Security}, vol.~6, no.~3, pp. 551--564, Sep. 2011.

\end{thebibliography}

\end{document}